\documentclass[twocolumn,prl]{revtex4}

\usepackage{dcolumn}
\usepackage{amsmath}

\usepackage{graphicx}

\newcommand{\e}{\mathrm{e}}
\newcommand{\half}{\mbox{$\frac12$}}
\newcommand{\set}[1]{\lbrace#1\rbrace}
\newcommand{\av}[1]{\langle#1\rangle}

\newcommand{\defn}{\textit}
\renewcommand{\vec}{\mathbf}
\newcommand{\mat}{\mathbf}

\begin{document}

\title{Random graphs with clustering}
\author{M. E. J. Newman}
\affiliation{Department of Physics,
University of Michigan, Ann Arbor, MI 48109}
\affiliation{Santa Fe Institute, 1399 Hyde Park Road, Santa Fe, NM 87501}
\begin{abstract}
  We offer a solution to a long-standing problem in the physics of
  networks, the creation of a plausible, solvable model of a network that
  displays clustering or transitivity---the propensity for two neighbors of
  a network node also to be neighbors of one another.  We show how standard
  random graph models can be generalized to incorporate clustering and give
  exact solutions for various properties of the resulting networks,
  including sizes of network components, size of the giant component if
  there is one, position of the phase transition at which the giant
  component forms, and position of the phase transition for percolation on
  the network.
\end{abstract}
\maketitle

Many networks, perhaps most, show clustering or transitivity, the
propensity for two neighbors of the same vertex also to be neighbors of one
another, forming a triangle of connections in the
network~\cite{Rapoport68,WS98,SB06a}.  In a social network of friendships
between individuals, for example, there is a high probability that two
friends of a given individual will also be friends of one another.  The
network average of this probability is called the \defn{clustering
  coefficient} for the network.  Measured clustering coefficients for
social networks are typically on the order of tens of percent and similar
values are seen in many nonsocial networks as well, including technological
and biological networks~\cite{NP03b}.

Although clustering in networks has been known and discussed for many
years, it has proved difficult to model mathematically.  Our continued
inability to create a plausible analytic model of clustered networks has
been a substantial impediment to the development of a comprehensive theory
of networked systems and contrasts sharply with our successes in the
modeling of other network properties such as degree
distributions~\cite{BA99b,NSW01} and correlations~\cite{PVV01,Newman02f}.
A few network models, such as the small-world model of Watts and
Strogatz~\cite{WS98}, do show clustering and are at least approximately
solvable, but are also rather specialized and not suitable as models of
most real networks.  A large number of computational models of clustered
networks have been proposed that are more general in scope, almost all
based on some form of ``triadic closure'' process in which one searches an
initially unclustered network for pairs of vertices with a common neighbor
and then connects them to form
triangles~\cite{JGN01,HK02b,KE02,SB05,BKM09}.  Unfortunately, because of
the nature of these models, the calculation of their properties is limited
to numerical approaches~\cite{note1}.

An ideal solution to these problems would be to generalize the standard
random graphs that form the foundation for much of modern network theory to
create an ensemble model of clustered networks for which one could
calculate ensemble average properties exactly.  It has long been felt,
however, that such an approach is likely to be unworkable because our
ability to calculate the properties of random graphs rests on the fact that
they are ``locally tree-like,'' i.e.,~that they contain no short loops in
their structure.  The triangles of clustered networks violate this
condition and hence one would expect their introduction into a random graph
model to render the model intractable.

But in this paper we show that this is not the case.  We show that it is in
fact possible to generalize random graphs to incorporate clustering in a
simple, sensible fashion and to derive exact formulas for a wide variety of
properties of the resulting networks.

\begin{figure}[b]
\begin{center}
\includegraphics[width=5.5cm]{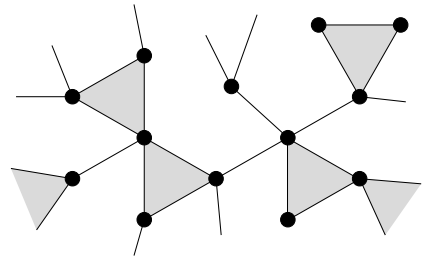}
\end{center}
\caption{In the model proposed here, we separately specify the number of
  single edges and complete triangles (shaded) attached to each vertex.}
\label{fig:model}
\end{figure}

The model we propose generalizes the standard ``configuration model'' of
network theory, which is a model of a random graph with arbitrary degree
distribution~\cite{MR95,NSW01}.  In that model one specifies the number of
edges connected to each vertex.  In our generalized model, pictured in
Fig.~\ref{fig:model}, we specify both the number of edges and the number of
triangles.  For a network of $n$ vertices, we define $t_i$ to be the number
of triangles in which vertex~$i$ participates and $s_i$ to be the number of
single edges other than those belonging to the triangles.  That is, edges
within triangles in this model are enumerated separately from edges that
are placed singly~\cite{note2}.  We can think of a single edge as being a
network element that joins together two vertices and a triangle as a
different kind of element that joins three.  In principle, one could
generalize the model further to include higher-order elements of four or
more vertices.  The techniques described here can be extended in a
straightforward fashion to such cases.

We can think of $s_i$ as specifying the number of ends or ``stubs'' of
single edges that emerge from vertex~$i$ and $t_i$ as specifying the number
of corners of triangles.  The complete joint degree sequence
$\set{s_i,t_i}$ specifies the numbers of such stubs and corners for every
vertex.  In simple cases the values of $s_i$ and $t_i$ may be uncorrelated,
but correlated choices are also possible that allow us to reproduce more
complex behaviors seen in some networks, such as variation of local
clustering with degree~\cite{Ravasz02}.

Given the degree sequence, we create our network by choosing pairs of stubs
uniformly at random and joining them to make complete edges, and also
choosing trios of corners at random and joining them to form complete
triangles.  The end result is a network drawn uniformly at random from the
set of all possible matchings of stubs and corners.  The only constraint is
that, in order that there be no stubs or corners left over at the end of
the process, the total number of stubs must be a multiple of~2 and the
total number of corners a multiple of~3.

We define the joint degree distribution~$p_{st}$ of our network to be the
fraction of vertices connected to $s$ single edges and $t$ triangles---a
quantity that can be easily measured for any observed network.  Given this
joint distribution, the conventional degree distribution of the network,
the probability~$p_k$ that a vertex has $k$ edges in total, both singly and
in triangles, is
\begin{equation}
p_k = \sum_{s,t=0}^\infty p_{st} \delta_{k,s+2t},
\end{equation}
since each triangle connected to a vertex contributes 2 to the degree and
each single edge contributes~1.  (Here $\delta_{ij}$ is the Kronecker
delta.)

As with other random graph models, calculations for the model presented in
this paper make use of probability generating functions.  The generating
function for the joint degree distribution of our network is a function of
two variables thus:
\begin{equation}
g_p(x,y) = \sum_{s,t=0}^\infty p_{st}\,x^s y^t.
\end{equation}
We can also write down a generating function for the total degree
distribution~$p_k$ thus:
\begin{align}
f(z) &= \sum_{k=0}^\infty p_k z^k
      = \sum_{k=0}^\infty \sum_{s,t=0}^\infty p_{st} \delta_{k,s+2t} z^k
      = \sum_{s,t=0}^\infty p_{st} z^{s+2t} \nonumber\\
     &= g_p(z,z^2).
\end{align}

We can use these generating functions to calculate, for instance, the
clustering coefficient~$C$ of the network.  The clustering coefficient can
be defined as~\cite{NSW01}
\begin{equation}
C = {3\times\mbox{(number of triangles in network)}\over
     \mbox{(number of connected triples)}}
  = {3N_\triangle\over N_3},
\label{eq:defsc}
\end{equation}
where a connected triple means a single vertex connected by edges to two
others.  For the present model we have
\begin{align}
3N_\triangle &= n\sum_{st} tp_{st}
              = n \biggl( {\partial g_p\over\partial y} \biggr)_{x=y=1}, \\
N_3 &= n \sum_k {k\choose2} p_k
     = \half n \biggl( {\partial^2 f\over\partial z^2} \biggr)_{z=1},
\end{align}
and substituting into~\eqref{eq:defsc} then gives us the value of the
clustering coefficient.  Note that the factors of $n$ cancel out in the
substitution, giving a value of $C$ that remains nonzero in the limit
$n\to\infty$ so that the network always has clustering, by contrast with
the configuration model and similar random graphs for which $C\to0$.

A further quantity that will be important in the following calculations is
the so-called \defn{excess degree distribution}~\cite{NSW01}.  In the
current model there are actually two different excess degree distributions:
\begin{equation}
q_{st} = {(s+1)p_{s+1,t}\over\av{s}},\qquad
r_{st} = {(t+1)p_{s,t+1}\over\av{t}},
\end{equation}
where $\av{s}$ and $\av{t}$ are the averages of $s$ and~$t$ over all
vertices.  Here $q_{st}$ is the distribution of the number of edges and
triangles attached to a vertex reached by traversing an edge, excluding the
traversed edge, and $r_{st}$ is the corresponding distribution for a vertex
reached by traversing a triangle.  The generating functions for these
distributions are
\begin{align}
\label{eq:defsgq}
g_q(x,y) &= \sum_{st} q_{st} x^s y^t
          = {1\over\av{s}} \sum_{st} s p_{st} x^{s-1} y^t
          = {1\over\av{s}} {\partial g_p\over\partial x}, \\
\label{eq:defsgr}
g_r(x,y) &= \sum_{st} r_{st} x^s y^t
          = {1\over\av{t}} \sum_{st} t p_{st} x^s y^{t-1}
          = {1\over\av{t}} {\partial g_p\over\partial y}.
\end{align}

One of the definitive features of any network is its giant component---the
portion of the network that is connected into a single extensive group such
that any vertex in the group can be reached from any other via the network.
In a communication network, for example, the giant component corresponds to
the fraction of vertices that can actually intercommunicate, the rest being
isolated in disconnected small components.  We can use our generating
functions to calculate the size of the giant component in the clustered
network.

Let $u$ be the mean probability that a vertex reached by traversing a
single edge is \emph{not} a member of the giant component and $v$ be the
corresponding probability for a vertex reached by traversing a triangle.
(Equivalently, $v^2$~is the probability that a triangle doesn't lead to the
giant component via either of the vertices at its other corners.)  In order
for a vertex at the end of a single edge not to belong to the giant
component, all the other vertices to which it is connected, either by edges
or by triangles, must also not be members of the giant component.  If it is
connected to $s$ other edges and $t$ triangles, then this happens with
probability $u^s v^{2t}$.  The generalized degrees~$s$ and $t$ are
distributed according to the excess degree distribution $q_{st}$ and,
averaging over this distribution, we find
\begin{equation}
u = \sum_{st} q_{st} u^s v^{2t} = g_q(u,v^2).
\label{eq:gcu}
\end{equation}
By a similar argument we also find that
\begin{equation}
v = g_r(u,v^2).
\label{eq:gcv}
\end{equation}
Then the probability that a randomly chosen vertex is not in the giant
component is $\sum_{st} p_{st} u^s v^{2t} = g_p(u,v^2)$ and the expected
size~$S$ of the giant component as a fraction of the entire network is one
minus this quantity:
\begin{equation}
S = 1 - g_p(u,v^2).
\label{eq:gcs}
\end{equation}
Between them, Eqs.~\eqref{eq:gcu}--\eqref{eq:gcs} allow us to calculate the
size of the giant component if there is one.

\begin{figure}
\begin{center}
\includegraphics[width=8cm]{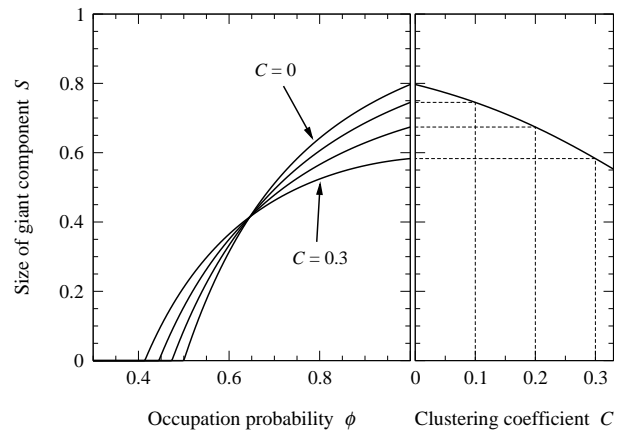}
\end{center}
\caption{ Right panel: the size of the giant component in networks with the
  degree distribution of Eq.~\eqref{eq:poisson} and average degree
  $\mu+2\nu=2$, as a function of clustering coefficient.  Left panel: the
  size of the giant cluster for percolation on the same networks for values
  of the clustering coefficient $C=0$, 0.1, 0.2, and~0.3, as a function of
  bond occupation probability~$\phi$.  Note that when $\phi=1$ the giant
  cluster and giant component have the same size, as indicated by the
  dotted lines.}
\label{fig:perc}
\end{figure}

As an example, consider a network that has the doubly Poisson degree
distribution
\begin{equation}
p_{st} = \e^{-\mu} {\mu^s\over s!}\,\e^{-\nu} {\nu^t\over t!},
\label{eq:poisson}
\end{equation}
where the parameters $\mu$ and $\nu$ are the average numbers of single
edges and triangles per vertex respectively.  Then
\begin{equation}
g_p(x,y) = g_q(x,y) = g_r(x,y) = \e^{\mu(x-1)}\e^{\nu(y-1)},
\end{equation}
and $u=v=1-S$, leading to
\begin{equation}
S = 1 - \e^{-[\mu S + \nu S(2-S)]}.
\end{equation}
This is a transcendental equation that has no closed-form solution (other
than the trivial solution $S=0$) but it can easily be solved by numerical
iteration starting from a suitable initial value.  The right-hand panel of
Fig.~\ref{fig:perc} shows the resulting giant component size as a function
of clustering coefficient for a network with fixed average degree.  As the
figure shows, the size of the giant component falls off with increasing
clustering coefficient, which happens because the triangles that give the
network its clustering contain redundant edges that serve no purpose in
connecting the giant component together.  One edge out of every three in a
triangle is redundant in this way.  Thus for a given average degree, and
hence a given total number of edges, fewer vertices can be connected
together in a network of triangles than in a network of single edges.

We can also calculate the sizes of the small components in the network.
Let $h_q(z)$ be the generating function for the distribution of number of
vertices accessible, either directly or indirectly, via the vertex at the
end of a single edge, and similarly for $h_r(z)$ and triangles.  Then, by
an argument analogous to that of~\cite{NSW01}, we can show that
\begin{equation}
h_q(z) = z g_q( h_q(z), h_r^2(z) ),\quad
h_r(z) = z g_r( h_q(z), h_r^2(z) ),
\label{eq:hqhr}
\end{equation}
and the probability that a randomly chosen vertex anywhere in the network
belongs to a component of a given size is generated by
\begin{equation}
h_p(z) = z g_p( h_q(z), h_r^2(z) ).
\end{equation}
Then, for example, the mean size of the component to which a vertex belongs
is
\begin{equation}
h_p'(1) = 1 + g_p^{(1,0)}(1,1) h_q'(1) + 2 g_p^{(0,1)}(1,1) h_r'(1),
\label{eq:avs}
\end{equation}
where $g_p^{(m,n)}$ is $g_p$ differentiated $m$ times with respect to its
first argument and $n$ times with respect to its second.

The derivatives $h_q'(1)$ and $h_r'(1)$ in Eq.~\eqref{eq:avs} can be found
from Eq.~\eqref{eq:hqhr} by differentiating, setting $x=y=1$, and making
use Eqs.~\eqref{eq:defsgq} and~\eqref{eq:defsgr}, which gives
\begin{align}
\label{eq:hqp}
h_q'(1) &= 1 + {1\over\av{s}} H_{xx} h_q'(1)
             + {2\over\av{s}} H_{xy} h_r'(1), \\
\label{eq:hrp}
h_r'(1) &= 1 + {1\over\av{t}} H_{yx} h_q'(1)
             + {2\over\av{t}} H_{yy} h_r'(1),
\end{align}
where the $H$ variables are the elements of the Hessian matrix~$\mat{H}$ of
second derivatives of $g_p$, evaluated at the point $x=y=1$:
\begin{equation}
H_{xx} = g_p^{(2,0)}(1,1),\qquad
H_{xy} = g_p^{(1,1)}(1,1),
\label{eq:hessian}
\end{equation}
and so forth.

We can write Eqs.~\eqref{eq:hqp} and~\eqref{eq:hrp} in matrix form as
$\vec{h} = \vec{1} +
\boldsymbol{\alpha}^{-1}\mat{H}\boldsymbol{\beta}\cdot\vec{h}$, where the
vectors~$\vec{h}$ and $\vec{1}$ are $\vec{h}=(h_q'(1),h_r'(1))$ and
$\vec{1}=(1,1)$ and the diagonal matrices $\boldsymbol{\alpha}$
and~$\boldsymbol{\beta}$ are
\begin{equation}
\boldsymbol{\alpha} = \begin{pmatrix} \av{s} & 0 \\ 0 & \av{t} \end{pmatrix},
\qquad
\boldsymbol{\beta} = \begin{pmatrix} 1 & 0 \\ 0 & 2 \end{pmatrix}.
\end{equation}
Rearranging yields $(\mat{I} -
\boldsymbol{\alpha}^{-1}\mat{H}\boldsymbol{\beta})\cdot\vec{h} = \vec{1}$,
where $\mat{I}$ is the identity matrix, and by inverting this equation and
combining the result with Eq.~\eqref{eq:avs} we can find the average
component size.

The average will diverge at the point where $\det(\mat{I} -
\boldsymbol{\alpha}^{-1}\mat{H}\boldsymbol{\beta}) = 0$ and, performing the
derivatives in Eq.~\eqref{eq:hessian}, we find the following condition for
the point at which the giant component forms:
\begin{equation}
\biggl[ {\av{s^2}\over\av{s}}-2 \biggr]
  \biggl[ 2{\av{t^2}\over\av{t}}-3 \biggr] = 2{\av{st}^2\over\av{s}\av{t}}.
\label{eq:mr}
\end{equation}
In the case where there are no triangles in the network, this equation
reduces to the well known criterion $\av{s^2}/\av{s}-2=0$ of Molloy and
Reed~\cite{MR95} for the phase transition in the ordinary configuration
model.  When triangles are present, Eq.~\eqref{eq:mr}~gives the appropriate
generalization of that criterion.

We can calculate many other properties of our networks, including average
path lengths and vertex connection probabilities.  As our final example in
this paper we demonstrate the calculation of percolation properties of
random graphs with clustering.  Both site and bond percolation processes on
networks have important applications: site percolation is related to
network resilience~\cite{CEBH00,CNSW00}, while bond percolation is related
to the dynamics of disease and other spreading
processes~\cite{Mollison77,Grassberger82}.  Consider, for instance, a bond
percolation process on our model network, with each edge in the network
occupied independently with probability~$\phi$.  By analogy with our
earlier calculations, let $u$ be the probability that a vertex is not
connected to the percolating (giant) cluster of this percolation process
via one of its single edges, and let $v^2$ be the corresponding probability
for a triangle.

If a vertex is not connected to the giant cluster via a given single edge
then one of two things must be true: either the edge is not occupied, which
happens with probability $1-\phi$, or it is occupied but the vertex at its
end is itself not connected to the giant cluster via any of its other edges
or triangles of which, let us say, there are $s$ and~$t$ respectively.
This second process happens with probability $\phi u^s v^{2t}$.  But $s$
and $t$ are by definition distributed according to the excess degree
distribution $q_{st}$ and, averaging over this distribution, we then find
that
\begin{equation}
u = 1 - \phi + \phi \sum_{st} q_{st} u^s v^{2t} = 1 - \phi [1-g_q(u,v^2)].
\label{eq:percu}
\end{equation}
The corresponding equation for triangles is more involved, but still
essentially straightforward to derive:
\begin{equation}
v^2 = 1 - 2\phi(1-\phi^2) [1-g_r(u,v^2)]
        - \phi^2(3-2\phi) [1-g_r^2(u,v^2)].
\label{eq:percv}
\end{equation}
(Notice that Eqs.~\eqref{eq:percu} and~\eqref{eq:percv} reduce to
Eqs.~\eqref{eq:gcu} and~\eqref{eq:gcv} for the giant component of the
network, as they should, when $\phi=1$.)

Now the size~$S$ of the giant cluster of the percolation process is given
by $S = 1 - g_p(u,v^2)$.  The left-hand panel of Fig.~\ref{fig:perc} shows
$S$ as a function of~$\phi$ for the Poisson network of
Eq.~\eqref{eq:poisson}, for fixed average degree and several different
values of the clustering coefficient.  As the figure shows, higher
clustering pushes the percolation transition toward lower values of~$\phi$,
which can be understood as an effect of the redundant paths introduced by
the triangles in the network, which provide more opportunities to connect
clusters together.  At the same time, the ultimate size of the giant
cluster as $\phi$ approaches~1 is smaller in more clustered networks and
indeed becomes equal to the size of the giant component when $\phi=1$, as
indicated by the dashed lines in the figure.  Other properties of the
percolation process can be calculated in a similar fashion, including the
position of the percolation threshold, the mean size of small clusters, and
the complete distribution of sizes of small clusters.

To conclude, we have proposed a random-graph model of a clustered network
that is exactly solvable for many of its properties including component
sizes, existence and size of a giant component, and percolation properties.
The model answers a long-standing question in the study of networks by
showing how to construct an unbiased ensemble of networks with clustering,
and could form the basis for future investigations of the effects of
clustering on many processes of interest, including epidemic processes,
network resilience, and dynamical systems on networks.

The author thanks Brian Karrer and Lenka Zdeborova for useful
conversations.  This work was funded in part by the National Science
Foundation under grant DMS--0804778.


\begin{thebibliography}{10}
\expandafter\ifx\csname url\endcsname\relax
  \def\url#1{\texttt{#1}}\fi
\expandafter\ifx\csname urlprefix\endcsname\relax\def\urlprefix{URL }\fi

\bibitem{Rapoport68}
A.~Rapoport, Cycle distribution in random nets. \textit{Bulletin of
  Mathematical Biophysics} \textbf{10}, 145--157 (1968).

\bibitem{WS98}
D.~J. Watts and S.~H. Strogatz, Collective dynamics of `small-world' networks.
  \textit{Nature} \textbf{393}, 440--442 (1998).

\bibitem{SB06a}
M.~A. Serrano and M.~Bogu{\~n}\'a, Clustering in complex networks: {I}.
  {G}eneral formalism. \textit{Phys. Rev. E} \textbf{74}, 056114 (2006).

\bibitem{NP03b}
M.~E.~J. Newman and J.~Park, Why social networks are different from other types
  of networks. \textit{Phys. Rev. E} \textbf{68}, 036122 (2003).

\bibitem{BA99b}
A.-L. Barab\'asi and R.~Albert, Emergence of scaling in random networks.
  \textit{Science} \textbf{286}, 509--512 (1999).

\bibitem{NSW01}
M.~E.~J. Newman, S.~H. Strogatz, and D.~J. Watts, Random graphs with arbitrary
  degree distributions and their applications. \textit{Phys. Rev. E}
  \textbf{64}, 026118 (2001).

\bibitem{PVV01}
R.~Pastor-Satorras, A.~V\'azquez, and A.~Vespignani, Dynamical and correlation
  properties of the {I}nternet. \textit{Phys. Rev. Lett.} \textbf{87}, 258701
  (2001).

\bibitem{Newman02f}
M.~E.~J. Newman, Assortative mixing in networks. \textit{Phys. Rev. Lett.}
  \textbf{89}, 208701 (2002).

\bibitem{JGN01}
E.~M. Jin, M.~Girvan, and M.~E.~J. Newman, The structure of growing social
  networks. \textit{Phys. Rev. E} \textbf{64}, 046132 (2001).

\bibitem{HK02b}
P.~Holme and B.~J. Kim, Growing scale-free networks with tunable clustering.
  \textit{Phys. Rev. E} \textbf{65}, 026107 (2002).

\bibitem{KE02}
K.~Klemm and V.~M. Eguiluz, Highly clustered scale-free networks. \textit{Phys.
  Rev. E} \textbf{65}, 036123 (2002).

\bibitem{SB05}
M.~A. Serrano and M.~Bogu{\~n}\'a, Tuning clustering in random networks with
  arbitrary degree distributions. \textit{Phys. Rev. E} \textbf{72}, 036133
  (2005).

\bibitem{BKM09}
S.~Bansal, S.~Khandelwal, and L.~A. Meyers, Evolving clustered random networks.
  Preprint arxiv:0808.0509 (2008).

\bibitem{note1}
  A recent preprint of B. Bollob\'as, S. Janson, and O.  Riordan
  [arxiv:0807.2040 (2008)] describes an interesting and general model
  capable of creating clustered networks among many other possibilities.
  Because of its complexity the model appears challenging to treat
  analytically and only a few results are known, but it is possible that
  for special cases calculations similar to those described in this paper
  could be carried out.

\bibitem{MR95}
M.~Molloy and B.~Reed, A critical point for random graphs with a given degree
  sequence. \textit{Random Structures and Algorithms} \textbf{6}, 161--179
  (1995).

\bibitem{note2}
  It is possible for single edges by chance to form triangles themselves,
  but it is straightforward to show that, so long as mean degree remains
  constant as~$n$ increases, the density of such triangles vanishes in the
  limit of large system size.  Similarly the density of multiedges---of two
  vertices being connected by two or more different edges or triangles or a
  combination of the two---vanishes in the limit of large~$n$.

\bibitem{Ravasz02}
E.~Ravasz, A.~L. Somera, D.~A. Mongru, Z.~Oltvai, and A.-L. Barab\'asi,
  Hierarchical organization of modularity in metabolic networks.
  \textit{Science} \textbf{297}, 1551--1555 (2002).

\bibitem{CEBH00}
R.~Cohen, K.~Erez, D.~{ben-Avraham}, and S.~Havlin, Resilience of the
  {I}nternet to random breakdowns. \textit{Phys. Rev. Lett.} \textbf{85},
  4626--4628 (2000).

\bibitem{CNSW00}
D.~S. Callaway, M.~E.~J. Newman, S.~H. Strogatz, and D.~J. Watts, Network
  robustness and fragility: Percolation on random graphs. \textit{Phys. Rev.
  Lett.} \textbf{85}, 5468--5471 (2000).

\bibitem{Mollison77}
D.~Mollison, Spatial contact models for ecological and epidemic spread.
  \textit{Journal of the Royal Statistical Society B} \textbf{39}, 283--326
  (1977).

\bibitem{Grassberger82}
P.~Grassberger, On the critical behavior of the general epidemic process and
  dynamical percolation. \textit{Math. Biosci.} \textbf{63}, 157--172 (1982).

\end{thebibliography}
\end{document}